\documentclass[a4paper,11pt]{article}
\usepackage{amsmath}
\usepackage{amsthm}
\usepackage{amssymb}
\usepackage{amsfonts}
\usepackage{amscd}
\usepackage[latin1]{inputenc}

\theoremstyle{theorem}
\newtheorem{theorem}{Theorem}

\newtheorem{corollary}[theorem]{Corollary}
\newtheorem{lemma}[theorem]{Lemma}

\theoremstyle{definition}

\newtheorem{assumption}{Assumption}

\theoremstyle{remark}
\newtheorem{remark}[theorem]{Remark}

\newcommand{\al}{\alpha}
\newcommand{\be}{\beta}

\newcommand{\ep}{\epsilon}

\newcommand{\ka}{\kappa}
\newcommand{\la}{\lambda}
\newcommand{\om}{\omega}
\newcommand{\si}{\sigma}

\newcommand{\vp}{\varphi}

\newcommand\Om\Omega
\newcommand\Te\Theta
\newcommand{\De}{\Delta}
\newcommand{\Ga}{\Gamma}
\newcommand{\La}{\Lambda}
\newcommand{\Si}{\Sigma}
\def\CC{\mathbb{C}}
\def\NN{\mathbb{N}}
\def\RR{\mathbb{R}}
\def\ZZ{\mathbb{Z}}

\newcommand{\cB}{{\mathcal B}}

\newcommand{\cD}{{\mathcal D}}

\newcommand{\cL}{{\mathcal L}}

\newcommand{\cN}{{\mathcal N}}
\newcommand{\cO}{{\mathcal O}}
\newcommand{\cP}{{\mathcal P}}

\newcommand{\cR}{{\mathcal R}}

\newcommand{\cT}{{\mathcal T}}

\newcommand{\pd}{\partial}
\newcommand\minus\backslash

\newcommand\lan\langle
\newcommand\ran\rangle

\newcommand{\I}{{\mathrm i}}
\newcommand{\e}{{\mathrm e}}

\DeclareMathOperator\lcm{lcm} \DeclareMathOperator\supp{supp}  \DeclareMathOperator\imag{Im}

\newcommand\AdS{\mathrm{AdS}}
\newcommand\Ric{\mathrm{Ric}}

\newcommand\cY{\mathcal Y}
\renewcommand\SS{\RR/2\pi\ZZ}
\newcommand\diff[1]{\frac{\pd}{\pd #1}}
\newcommand\tT{\widetilde T}
\newcommand\tw{\widetilde w}
\newcommand\tv{\widetilde v}
\newcommand\hv{\hat v}
\newcommand\tf{\widetilde f}
\newcommand\cV{\mathcal V}
\newcommand\hf{\hat f}
\newcommand\tL{\widetilde L}
\newcommand\tvp{\widetilde \vp}
\renewcommand\geq\geqslant
\renewcommand\leq\leqslant
\newcommand\tcL{\widetilde{\mathcal L}}
\newcommand\AC{\mathcal{AC}^1} %\mathit{AC}^1}
\title{\Large\bf Global causal propagator for the Klein--Gordon equation\\ on a class of  supersymmetric AdS backgrounds}
\author{Alberto Enciso$^a$\thanks{alberto.enciso@math.ethz.ch}\;
  and Niky Kamran$^b$\thanks{nkamran@math.mcgill.ca}}
\date{\small $^a$ Departement Mathematik, ETH Zürich, 8092 Zürich, Switzerland.\vspace{1ex}\\
$^b$ Department of Mathematics and Statistics, McGill University, Montréal, Québec,\\ Canada H3A 2K6.}

\begin{document}
\maketitle

\begin{abstract}
We analyze the Cauchy problem for the Klein--Gordon equation on the
type IIB supergravity backgrounds $\AdS^5× Y^{p,q}$, where $Y^{p,q}$
is any of the Sasaki--Einstein $5$-manifolds introduced by
Gaunlett, Martelli, Sparks and Waldram (Adv.\ Theor.\ Math.\ Phys.\
8 (2004) 711--734). Although these spaces are not globally
hyperbolic, we prove that there exists a unique admissible
propagator and derive an integral representation thereof using
spectral-theoretic techniques.
\end{abstract}

\section{Introduction}

Sasaki--Einstein manifolds, that is Einstein manifolds whose metric
cone is Calabi--Yau, are of considerable interest in Physics because
of their connections to the AdS/CFT correspondence~\cite{AGM00}.
More precisely, if $Y$ is a Sasaki--Einstein $5$-manifold, then the
product manifold $\AdS^5× Y$ is a solution of type IIB supergravity
that is conjectured to be dual to an $\cN=1$ superconformal field
theory in four dimensions. In particular, the AdS/CFT correspondence
implies that the asymptotic behavior at infinity of the
Klein--Gordon propagator will carry information on the correlation
functions of the dual superconformal field theory in four
dimensions~\cite{AGM00}.

The first explicit examples of Sasaki--Einstein geometries in five
dimensions (other than the round $5$-sphere, the homogeneous metric
$T^{1,1}$ on $S^2× S^3$ and some quotients thereof) were discovered
by Gauntlett, Martelli, Sparks and Waldram in~\cite{GMSW04a}. These
manifolds, labeled by pairs of integers $(p,q)$ and denoted by
$Y^{p,q}$, are constructed as $S^1$-bundles over an axially squashed
$S^2$-bundle over the $2$-sphere. A description of the spaces
$Y^{p,q}$ as cohomogeneity-$1$ manifolds has been given by
Conti~\cite{Co07}, while examples of Sasaki--Einstein manifolds in
higher dimensions have been obtained in~\cite{GMSW04b,CL05}. The
associated family of supergravity solutions $\AdS^5× Y^{p,q}$
includes both quasi-regular manifolds~\cite{BG00}, which are dual to
superconformal field theories with compact $R$-symmetry and rational
central charges, and irregular manifolds, dual to field theories
with noncompact $R$-symmetry and irrational central charges. A
detailed construction of the dual superconformal quiver gauge
theories was presented in~\cite{JHEP}.

In this paper we aim to rigorously analyze the Cauchy problem for
the Klein--Gordon equation on the manifolds $\AdS^5× Y^{p,q}$
through the construction of a global causal propagator. Although the
spaces $\AdS^5× Y^{p,q}$ are not globally hyperbolic, we will see
that under mild technical assumptions a unique propagator exists and can be
constructed explicitly through spectral methods. Our main result
(Theorem~\ref{T.main} and Corollary~\ref{C.inhom}) is a spectral
integral representation for this propagator.

The approach we take exploits the separability of the $\AdS^5×
Y^{p,q}$ metrics to compute the eigenfunctions of the Laplace
operator in $Y^{p,q}$ in quasi closed form, by expressing them in
terms of the eigenfunctions of the Friedrichs extension of a single
second-order ordinary differential operator with seven regular
singular points. The subtle geometry of the spaces $Y^{p,q}$
introduces additional complications in the analysis, since the
`angular' variables in which the metric of $Y^{p,q}$ separates are
not defined globally. In order to circumvent this problem, we start
by constructing a Fourier-type decomposition of the space of
square-integrable functions on $Y^{p,q}$ adapted to the global
structure of the manifold and to the action of the Laplacian. Once
the eigenfunctions of the Laplacian in $Y^{p,q}$ have been computed,
the analysis of the Klein--Gordon equation in $\AdS^5× Y^{p,q}$ can
be reduced to that of a family of linear hyperbolic equations in
anti-de Sitter space. We discuss the existence and uniqueness of
causal propagators for these equations using Ishibashi and Wald's
spectral-theoretic approach to wave equations on static
space-times~\cite{Wa80,IW03,IW04}. For our purpose, this presents
several  advantages over the classical method of Riesz transforms,
since the latter method only yields local solutions to the Cauchy
problem in the case in which the underlying space-time is not
globally hyperbolic~\cite{BGP07}.

Our paper is organized as follows. In Section~\ref{S.Fourier} we
introduce a Fourier-type decomposition of $L^2(Y^{p,q})$
(Lemma~\ref{L.Y}) which is crucial to the rest of the paper. In
Section~\ref{S.Laplacian} we use this Fourier decomposition to
compute the eigenfunctions of the Laplacian in $Y^{p,q}$
(Theorem~\ref{T.spec}). The expression of these eigenfunctions is
totally explicit and involves the spectral decomposition of a single
ordinary differential operator, which we analyze in Lemma~\ref{L.S}.
Finally, in Section~\ref{S.wave} we prove that there exists a unique
physically admissible propagator for the Klein--Gordon equation in
$\AdS^5× Y^{p,q}$ and derive an integral representation thereof for
both the homogeneous and inhomogeneous Cauchy problem
(Theorem~\ref{T.main} and Corollary~\ref{C.inhom}). The article
concludes with an Appendix where we recall the conditions that a
propagator of a linear wave equation must satisfy in order to be
physically admissible.

\section{Fourier-type expansions in $Y^{p,q}$}
\label{S.Fourier}

After recalling some geometric facts about the family of
cohomogeneity-$1$ Sasaki--Einstein $5$-manifolds $Y^{p,q}$ recently
discovered in~\cite{GMSW04a}, our goal for this section is to introduce
a Fourier-type decomposition of the space of square-integrable
functions on $Y^{p,q}$ which is adapted to the geometry of the
manifold. This decomposition will be of crucial importance in the
rest of the article. Each manifold $Y^{p,q}$, labeled by two
positive integers $p<q<2p$, is an $S^1$-bundle over an axially
squashed $S^2$-bundle $B$ over a round $2$-sphere. It should be
noted that the integers $p$ and $q$ are not exactly the same as the
ones labeling the spaces $Y^{p,q}$ in~\cite{GMSW04a}; passing from
one set of integers to the other is straightforward, but for our
purposes it is slightly more convenient to define the labeling
integers as we will do below. We will recall several results on the
global geometry of the spaces $Y^{p,q}$ without further mention as
we need them; proofs of these facts and further discussion can be
found in~\cite{GMSW04a,MS06,MS09}.

We begin our discussion with the four-dimensional sphere bundle $B$
over $S^2$. We start with the local metric
\begin{equation}\label{gB}
  g_B:=\frac{dy^2}{w(y)\,r(y)}+ \frac{r(y)}9 \big(d\psi-\cos\theta\, d\phi\big)^2 + \frac{1-y}6\big(d\theta^2+\sin^2\theta\,d\phi^2\big)\,,
\end{equation}
where
\begin{equation}\label{wq}
  w(y):=\frac{2(a-y^2)}{1-y}\,,\qquad r(y):=\frac{a-3y^2+2y^3}{a-y^2}
\end{equation}
and $0<a<1$ is a real constant. The Riemannian volume measure
associated to $g_B$ is thus given by
\begin{equation}\label{dmuB}
d\mu_B:=\rho_B(y)\,dy\,\sin\theta\,d\theta\,d\phi\,d\psi\,,
\end{equation}
with
\[
\rho_B(y):=\frac{1-y}{18\,w(y)^{1/2}}\,.
\]
It was shown in~\cite{GMSW04a} that the above local metric defines a
unique complete 2-sphere bundle $B$ over $S^2$, which is conformally
Kähler and diffeomorphic to $S^2× S^2$.

In the following lemma we present a Fourier-type decomposition of
the space of $L^2$ functions on $B$ that is adapted to the above
coordinate system and which will be used in turn to give a Fourier
decomposition for the space of $L^2$ functions on $Y^{p,q}$. An
explicit description of the bundle structure of $B$ is required in
order to obtain the desired decomposition, so in the proof of the
lemma we indicate how $B$ is defined globally as a complete
manifold.

In order to state this lemma we set some notation. The cubic
polynomial
\begin{equation}\label{cubic}
a-3y^2+2y^3=0
\end{equation}
has three real roots for any $a\in (0,1)$, one negative and two
positive. In what follows, the negative root will be denoted by
$y_-$ and the smallest positive root by $y_+$ so that $y_-<0<y_+<a$.
We will also use the cover $\{V_1,V_2\}$ of $S^2$ given by
\begin{align*}%\label{V}
  V_1:=\big\{\theta\in[0,\pi),\;\phi\in\SS\big\}\,,\qquad
  V_2:=\big\{\theta\in(0,\pi],\;\phi\in\SS\big\}\,,
\end{align*}
so that $V_1$ (resp.\ $V_2$) stands for the sphere minus the north
(resp.\ south) pole.

\begin{lemma}\label{L.B}
  Let $\cB$ denote the complex Hilbert space
\[
\cB:=\Big\{\big(u_{nm}\big)_{n,m\in\ZZ}:
u_{nm}\in L^2\big((y_-,y_+),\rho_B(y)\,dy\big)\otimes L^2\big((0,\pi),\sin\theta\,d\theta\big)\Big\}\,,
\]
endowed with the norm defined by
\[
\Big\|\big(\Phi_{nm}\otimes\Theta_{nm}\big)_{n,m\in\ZZ}\Big\|_{\cB}^2:=\sum_{n,m\in
  \ZZ}\bigg(\int_{y_-}^{y_+} \big|\Phi_{nm}(y)\big|^2\,\rho_B(y)
\,dy\bigg)\bigg(\int_0^\pi\big|\Theta_{nm}(\theta)\big|^2 \sin\theta \,d\theta\bigg)\,.
\]
Then the map defined by
\begin{equation}\label{B}
\cB\ni\big(\Phi_{nm}\otimes\Theta_{nm}\big)_{n,m\in\ZZ}\mapsto
\sum_{n,m\in\ZZ}
\Phi_{nm}(y)\,\Theta_{nm}(\theta)\,\frac{\e^{\I(n\phi+2m\psi)}}{2\pi}\in
L^2(B)\,,
\end{equation}
defines an isomorphism between $\cB$ and $L^2(B)$. \end{lemma}
\begin{proof}
In order to show that the local metric~\eqref{gB} can be promoted to
a complete metric on a four-manifold, we choose $\theta\in[0,\pi]$
and $\phi\in\RR/ 2\pi\ZZ$ so that, for each fixed $y\in (y_-,y_+)$,
the last two terms in~\eqref{gB} yield the metric of a round
$2$-sphere.

The range of $y$ is taken to be $[y_-,y_+]$. This ensures that $w$
is strictly positive in this interval and $r\geq0$, vanishing only
at the endpoints $y_±$. If we identify $\psi$ periodically, the part
of $g_B$ given by
  \[
\frac{dy^2}{w(y)\,r(y)}+ \frac{r(y)}9d\psi^2
\]
describes a circle fibered over the interval $(y_-,y_+)$, the size
ofthe circle shrinking to zero at the endpoints. Remarkably, the
$(y,\psi)$ fibers are free of conical singularities if the period of
$\psi$ is $2\pi$, in which case the circles collapse smoothly and
the $(y,\psi)$ fibers are diffeomorphic to a $2$-sphere.

One must now check that the $2$-spheres described by the coordinates
$(y,\psi)$ fiber properly over the $2$-spheres defined by
$(\theta,\phi)$. For that, it suffices to consider the circles
associated to the coordinate $\psi$ for fixed $y\in (y_-,y_+)$, so
we will consider the corresponding $S^1$-bundles $B_y$ over $S^2$.
The curvature of the connection form $\frac1{2\pi}\cos\theta\,d\phi$
defines an integral de Rham cohomology class in the base because
\begin{equation}
-\frac1{2\pi}\int_{S^2}d\big(\cos\theta\,d\phi\big)=\frac1{2\pi}\int_0^{2\pi}\!\!\int_0^\pi \sin\theta\,d\theta\,d\phi=2\,.
\end{equation}
Therefore, for all $y$, a well-known theorem of
Kobayashi~\cite{Ko56} then ensures that $-\cos\theta\,d\phi$ defines
a connection on a principal $S^1$-bundle $B_y$ over $S^2$ isomorphic
to $S^3/\ZZ_2$ (or to the 2-sphere's unit tangent bundle). Since
$\{V_1,V_2\}$ is a trivializing cover of $S^2$, the $S^1$-bundle
$B_y$ is uniquely determined by this cover and the transition
function $F_{12}:V_1\cap V_2 \to\SS$ given by $F_{12}:=\pi$.

The set $V_1\cap V_2$ is the sphere minus the north and south poles,
so it has full measure in $V_1$ and $V_2$. Hence by the definition
of the transition function and the expression of the induced volume
element in $B_y$, an $L^2$ function on $B_y$ can be identified with
a measurable function $f^y:(0,\pi)×(\SS)× (\SS)\to\CC$ which
satisfies
\[
f^y(\theta,\phi,\psi)=f^y\big(\theta,\phi,\psi+F_{12}(\theta,\phi)\big)=f^y(\theta,\phi,\psi+\pi)
\]
a.e.\ and
\[
\int_0^{2\pi}\!\! \int_0^{2\pi}\!\!\int_0^{\pi}\big|f^y(\theta,\phi,\psi)\big|^2\,\sin\theta\,d\theta\,d\phi\,d\psi<\infty
\]
As usual, $\psi+\pi$ refers to the group operation in $\SS$, so it is to be understood modulo $2\pi$. This leads to the Fourier-type expansion
\[
f^y(\theta,\phi,\psi)=\sum_{n,m\in\ZZ}f^y_{nm}(\theta)\,\e^{\I(n\phi+2m\psi)}\,,
\]
which converges in $L^2$ sense. Formula~\eqref{B} immediately
follows from the latter equation by taking into account the
dependence on $y$ and using~\eqref{dmuB} and Fubini's theorem to
carry out the integration in $\phi$ and $\psi$.
\end{proof}

\begin{remark}
Since $\phi$ and $\psi$ have period $2\pi$, the difference between
Eq.~\eqref{B} and the ordinary Fourier decomposition for
$2\pi$-periodic functions of $\phi$ and $\psi$ lies in the fact that
only the even Fourier modes in $\psi$ appear in~\eqref{B}. As we
have seen, this reflects the fact that the bundle $B$ is diffeomorphic
but not isometric to the product of two round $2$-spheres. Notice that $F_{12}$ must be constant because
$\pd/\pd\psi$ is a globally defined Killing vector.
\end{remark}

We now consider Fourier decompositions of $L^2$ functions on the
manifolds $Y^{p,q}$. We begin with the local metric given by
\begin{equation}\label{gY}
  g:=g_B+w(y)\,\Big(d\al+ h(y)\,\big(d\psi -\cos\theta\,d\phi\big)\Big)^2\,,
\end{equation}
where
\begin{equation*}%\label{h}
  h(y):=\frac{a-2y+y^2}{6(a-y^2)}
\end{equation*}
and $g_B$ and $w$ are defined as in~\eqref{gB}--\eqref{wq}. It can
be verified that $g$ is (locally) Sasaki--Einstein, with $\Ric=4g$,
and that the corresponding Riemannian measure is
\begin{equation}\label{dmuY}
  d\mu:=\rho(y)\,\sin\theta\,dy\,d\theta\,d\phi\,d\psi\,d\al\,,
\end{equation}
whence $\rho(y):=(1-y)/18$.

It was proved in~\cite{GMSW04a} that, for any pair of positive
integers $p$ and $q$ with $p<q<2p$, one can choose a constant
$a\in(0,1)$ such that
\begin{equation}\label{pq}
  \frac{h(y_+)-h(y_-)}{2\,h(y_+)}=\frac pq\,,
\end{equation}
and that in this case there exists a unique complete, simply
connected manifold $Y^{p,q}$ whose metric is locally given
by~\eqref{gY}. Furthermore, $Y^{p,q}$ is a Sasaki--Einstein
principal $S^1$-bundle over $B$ diffeomorphic to $S^2× S^3$. In the
rest of the paper, we shall always assume that $a$ has been chosen
so that~\eqref{pq} is satisfied, with $p$ and $q$ coprime. Note that
\[
h(y_±)=\frac{y_±-1}{6y_±}\,,
\]
so $\mp h(y_±)>0$.

The main result of this section is the following lemma, where we
present a Fourier-type decomposition of $L^2(Y^{p,q})$ which will be
crucial for the rest of the paper. In order establish this
decomposition, we introduce another open cover $\{U_-,U_+\}$ of the
sphere, which will be used to describe the fibers of the sphere
bundle $B$. These sets again correspond to the whole sphere minus a
pole, and can be characterized in terms of $y$ and $\psi$ as
\begin{align*}%\label{U}
  U_-:=\big\{y\in[y_-,y_+),\;\psi\in\SS\big\}\,,\qquad  U_+:=\big\{y\in(y_-,y_+],\;\psi\in\SS\big\}\,.
\end{align*}
Moreover, let us call $\Si_0$ the $S^2$-fiber at any fixed point
$(\theta_0,\phi_0)$ in the base and let $\Si_±\simeq S^2$ be the submanifolds of $B$ given by $y=y_±$. It is then easy to see that
$\{\Si_0,\Si_+\}$ defines a basis of the homology group
$H_2(B;\ZZ)$. In the following lemma, we denote by
$\lcm\{x_1,\dots,x_k\}$ the least common multiple of the positive
integers $x_1,\dots,x_k$.

\begin{lemma}\label{L.Y}
  Let $\cY$ be the complex Hilbert space
\[
\cY:=\Big\{\big(u_{nml}\big)_{n,m,l\in\ZZ}:
u_{nml}\in L^2\big((y_-,y_+),\rho(y)\,dy\big)\otimes L^2\big((0,\pi),\sin\theta\,d\theta\big)\Big\}\,,
\]
endowed with the norm
\[
\Big\|\big(\Phi_{nml}\otimes\Theta_{nml}\big)_{n,m,l\in\ZZ}\Big\|_{\cY}^2:=\sum_{n,m,l\in
  \ZZ}\bigg(\int_{y_-}^{y_+} \big|\Phi_{nml}(y)\big|^2 \rho(y)
\,dy\bigg)\bigg(\int_0^{2\pi}\big|\Theta_{nml}(\theta)\big|^2 \sin\theta \,d\theta\bigg)\,,
\]
and let us set
\begin{equation}\label{tausi}
  \tau:=-2\,h(y_+)/q\,,\qquad \si:=\lcm\{2,pq,2p-q\}\,.
\end{equation}
Then the map defined by
\begin{equation}\label{Y}
\cY\ni\big(\Phi_{nml}\otimes\Theta_{nml}\big)_{n,m,l\in\ZZ}\mapsto
\sum_{n,m,l\in\ZZ}
\Phi_{nml}(y)\,\Theta_{nml}(\theta)\,\frac{\e^{\I(n\phi+2m\psi+\si
l\al/\tau)}}{(2\pi)^{3/2}}\in L^2(Y^{p,q})\,,
\end{equation}
defines an isomorphism between $\cY$ and $L^2(Y^{p,q})$.
\end{lemma}
\begin{proof}
If we periodically identify $\al$ by making it take values in
$\RR/2\pi\tau\ZZ$, then for each fixed $y\in[y_-,y_+]$ the term
$w(y)\,d\al^2$ in~\eqref{gY} describes a circle whose size does not
shrink to zero. To see that the metric~\eqref{gY} corresponds to a
complete compact manifold, notice that, by a theorem of
Kobayashi~\cite{Ko56}, $A:=h(y)(d\psi-\cos\theta\,d\phi)$ defines a
connection in a principal $S^1$-bundle over $B$ if and only if the
(globally defined) curvature $2$-form $dA/(2\pi\tau)$ defines an
integral de Rham cohomology class in $B$.

An easy computation shows that
  \[
\int_{\Si_0}\frac{dA}{2\pi\tau}=\frac{h(y_-)-h(y_+)}\tau\,,\qquad \int_{\Si_+}\frac{dA}{2\pi\tau}=\frac{2\,h(y_+)}\tau\,,
\]
By~\eqref{pq}, it follows that $dA/(2\pi\tau)\in H^2(B;\ZZ)$ if we
set $\tau:=-2\,h(y_+)/q$, in which case the periods of
$dA/(2\pi\tau)$ around $\Si_0$ and $\Si_+$ are respectively given by
$p$ and $-q$. Since $p$ and $q$ are coprime, it follows that
$Y^{p,q}$ is simply connected.

The bundle $Y^{p,q}$ is completely determined by the cover $\{V_i×
U_\ep: i=1,2,\; \ep=±\}$ and the associated transition functions
\[
F_{ij\ep\eta}:(V_i\cap V_j)\cap (U_\ep\cap U_\eta)\to \RR/2\pi\tau\ZZ\,.
\]
The transition functions $F_{ii-+}$, $i=1,2$, are easily derived using the fact that
\[
\int_{\Si_0}\frac{dA}{2\pi\tau}=p
\]
for any $(\theta_0,\phi_0)$, since this means that $d\al+ A$ defines
on $\Si_0$ an $S^1$-bundle with winding number $p$. Hence this
bundle is isomorphic to the lens space $S^3/\ZZ_p$~\cite{St51} and
$F_{ii-+}=2\pi\tau/p$. Similarly, $F_{12++}=-2\pi\tau/q$ because
\[
\int_{\Si_+}\frac{dA}{2\pi\tau}=-q\,.
\]

To determine $F_{12--}$, it suffices to observe that
\[
\int_{\Si_-}\frac{dA}{2\pi\tau}=2p-q
\]
by~\eqref{pq}. Notice that $2p-q$ is granted to be strictly
positive. In this case, $d\al+ A$ determines a connection on a
principal $S^1$-bundle over $\Si_-$ with winding number $2p-q$, so
that $F_{12--}=2\pi\tau/(2p-q)$. As
\[
F_{ij\ep\eta}= F_{ii\ep\eta}+F_{ij\eta\eta}\,,
\]
the full set of transition functions is uniquely determined from $F_{ii-+}$ and $F_{12\ep\ep}$.

Since $(V_i\cap V_j)× (U_\ep\cap U_\eta)$ has full measure in $V_i×
U_\ep$, an $L^2$ function in $Y^{p,q}$ can now be identified with a
measurable function
\[
f:(0,\pi)×(\SS)×(y_-,y_+)× (\SS)× (\RR/2\pi\tau\ZZ)\to \CC
\]
such that:
\begin{enumerate}
\item $f(\theta,\phi,y,\psi,\al)= f(\theta,\phi,y,\psi,\al+2\pi\tau/p) =f(\theta,\phi,y,\psi,\al-2\pi\tau/q) =f(\theta,\phi,y,\psi,\al+2\pi\tau/(2p-q))$ a.e.,
by definition of the transition functions $F_{ij\ep\eta}$.
\item $f(\theta,\phi,y,\psi,\al)= f(\theta,\phi,y,\psi+\pi,\al)=f(\theta,\phi,\psi,\al+\pi)$ a.e., because of the way the sets $V_i× U_\ep$
are patched to yield the bundle $B$, as analyzed in Lemma~\ref{L.B} using the auxiliary $S^1$-bundles $B_y$.
\item The integral
  \[
\int \big|f(\theta,\phi,y,\psi,\al)\big|^2\,\rho(y)\,dy\,\sin\theta\,d\theta\,d\phi\,d\psi\,d\al
\]
is finite, by the expression of the Riemannian measure~\eqref{dmuY}
\end{enumerate}
From~(i) and~(ii), we see that $f$ must be $2\pi\tau/\si$-periodic
in its last argument, with $\si:=\lcm\{pq,2p-q,2\}$ the least common
multiple of $2$, $p$, $q$ and $2p-q$, and $\pi$-periodic in its
fourth argument. This leads to the $L^2$ Fourier expansion
\[
f(\theta,\phi,y,\psi,\al) = \sum_{n,m,l\in\ZZ}f_{nml}(y,\theta)\,\e^{\I(n\phi+2m\psi+l\si\al/\tau)}\,,
\]
which readily gives~\eqref{Y} after recalling Eq.~\eqref{dmuY} and carrying out the integrals in $\phi$, $\psi$ and $\al$.
It should be noticed that all the Fourier frequencies compatible with the above periodicity condition must appear in the
decomposition formula due to the simple connectedness of $Y^{p,q}$.
\end{proof}

\section{The Laplacian in $Y^{p,q}$}
\label{S.Laplacian}

Our goal in this section is to derive a manageable formula for the
spectral resolution associated to the Laplacian in $Y^{p,q}$. As we
shall see, the computation of the spectral decomposition of the
Laplacian actually boils down to the analysis of a single Fuchsian
ordinary differential operator depending on three parameters.

It is well known that the Laplacian on $Y^{p,q}$, which we denote by
$\De$, defines a nonnegative, self-adjoint operator whose domain is
the Sobolev space $H^2(Y^{p,q})$ of square-integrable functions with
square-integrable second derivatives. The Laplacian is given in
local coordinates as
\begin{multline*}
  \De:=g^{ij}\nabla_i\nabla_j=\frac1{\rho(y)}\diff{y}\rho(y)\,w(y)\,r(y)\,\diff y +\frac1{w(y)}\frac{\pd^2}{\pd\al^2}+ \frac9{r(y)}\bigg(\diff\psi-h(y)\,\diff\al\bigg)^2\\
   +\frac6{1-y}\Bigg[\frac1{\sin\theta}\diff\theta\sin\theta\diff\theta +\frac1{\sin^2\theta}\bigg(\diff\phi+\cos\theta\diff\psi\bigg)^2\Bigg]\,.
\end{multline*}
Therefore, it is not difficult to see that the action of the
Laplacian on a function of the form
$u(y,\theta)\,\e^{\I(n\phi+2m\psi+l\si\al/\tau)}$ (which are
globally well defined, as discussed in Lemma~\ref{L.Y}) is given by
\begin{equation}\label{DeDenml} \De\big(u(y,\theta)\,\e^{\I(n\phi+2m\psi+l\si\al/\tau)}\big)=\big(\De_{nml}u(y,\theta)\big)\,\e^{\I(n\phi+2m\psi+l\si\al/\tau)}\,,
\end{equation}
where
\begin{equation}\label{Denml}
\De_{nml}:=\frac1{\rho(y)}\diff{y}\rho(y)\,w(y)\,r(y)\,\diff y -\frac1{w(y)}\bigg(\frac{\si l}\tau\bigg)^2- \frac9{r(y)}\bigg(2m-h(y)\,\frac{\si l}\tau\bigg)^2
+\frac6{1-y} T_{nm}
\end{equation}
and
\begin{equation}\label{Tnm}
  T_{nm}:=\frac1{\sin\theta}\diff\theta\sin\theta\diff\theta -\bigg(\frac{n+2m\cos\theta}{\sin\theta}\bigg)^2\,.
\end{equation}

Eq.~\eqref{DeDenml} suggests that the computation the spectral
resolution of $\De$ should be equivalent to finding an appropriate
orthonormal basis of eigenfunctions of the differential operators
$\De_{nml}$ on $L^2((y_-,y_+),\rho(y)\,dy)\otimes
L^2((0,\pi),\sin\theta\,d\theta)$. In the rest of this section we
shall show how this can be accomplished. We begin by constructing an
orthonormal basis of $L^2((0,\pi),\sin\theta\,d\theta)$ consisting
of eigenfunctions of $T_{nm}$. If $I\subset\RR$ is an open interval,
we shall denote by $\AC(I)$ the set of continuously differentiable
functions $v:I\to\CC$ whose derivative is absolutely continuous on
any compact subset of $I$.

\begin{lemma}\label{L.T}
Let us denote the Jacobi polynomials by $P^{(\bar a,\bar b)}_j$ and set
  \[
C_{nmj}:=\bigg(\frac{(2j+|n+2m|+|n-2m|+1)\,j!\,(j+|n+2m|+|n-2m|)!}{2\,(j+|n+2m|)!\,(j+|n-2m|)!}\bigg)^{1/2}\,.
\]
Then the analytic functions $v_{nmj}:[0,\pi]\to\RR$ given by
  \begin{equation*}
    v_{nmj}(\theta):=  C_{nmj}\, \sin^{|n+2m|}\tfrac\theta2\,\cos^{|n-2m|}\tfrac\theta2\, P_j^{(|n+2m|,|n-2m|)}(\cos\theta)\,,\qquad j\in\NN,
  \end{equation*}
define an orthonormal basis of $L^2((0,\pi),\sin\theta\,d\theta)$ and satisfy the eigenvalue equation $T_{nm}v_{nmj} =-\La_{nmj}v_{nmj}$, with
\[
\La_{nmj}:=2\Big(2j(j+1)+\big(|n+2m|+|n-2m|\big)(2j+1)+|n+2m||n-2m|+2m^2+n^2\Big)\,.
\]
\end{lemma}
\begin{proof}
In terms of the variable $z:=\sin^2\frac\theta2$, the differential equation $T_{nm}v(\theta)=-\La v(\theta)$
can be rewritten as
\begin{equation}\label{eqT2}
  \tT_{nm}\tv(z)=-\La \tv(z)\,,
\end{equation}
where
\begin{equation}\label{tT}
\tT_{nm}:=  z(1-z)\,\frac{\pd^2}{\pd z^2}+(1-2z)\, \diff z-\frac{(n+2m-4mz)^2}{4z(1-z)}
\end{equation}
and $\tv(z)$ stands for the expression of the function $v(\theta)$
in the variable $z$. It is not difficult to show that this equation
has three regular singular points, located at $0$, $1$ and $\infty$,
whose characteristic exponents are respectively given by $±(\frac
n2+m)$, $±(\frac n2-m)$ and $\frac12[1±(1+\La+4m^2)^{1/2}]$.

It then follows that the symmetric operator on $L^2((0,1))$ defined
by the action of~\eqref{tT} on $C^\infty_0((0,1))$ is in the limit
point case at $0$ (resp.\ at $1$) if and only if $n\neq -2m$ (resp.\
$n\neq 2m$). When both conditions are satisfied, there exists a
unique self-adjoint extension, whose domain is the set of $\tv\in
\AC((0,1))$ such that $\tT_{nm} \tv\in L^2((0,1))$. When $n\neq -2m$
(resp.\ $n\neq 2m$), we take the Friedrichs extension of the above
operator, which is determined~\cite{MZ00} by the boundary
condition(s)
\[
\lim_{z\searrow0}z\,\tv'(z)=0\qquad \Big(\text{resp. }\;\lim_{z\nearrow1}z\,\tv'(z)=0\Big)\,.
\]
We shall see that his choice of boundary condition(s) will preclude
the appearance of logarithmic singularities. With a slight abuse of
notation, we shall still denote by $\tT_{nm}$ the self-adjoint
operators under consideration.

If Eq.~\eqref{eqT2} holds, a simple calculation shows that
\[
\hv(z):=z^{-|n+2m|/2}\,(1-z)^{-|n-2m|/2}\,\tv(z)
\]
satisfies the hypergeometric equation
\begin{equation}\label{eqtw}
z(1-z)\,\hv''(z)+\big(\bar c-(\bar a+\bar b+1)z\big)\,\hv'(z)-\bar a\bar b\,\hv(z)=0\,,
\end{equation}
with parameters
\begin{gather*}
  2\bar a=1+|n+2m|+|n-2m|-\big(\La+4m^2+1\big)^{1/2}\,,\\ 2\bar b=1+|n+2m|+|n-2m|+\big(\La+4m^2+1\big)^{1/2}\,,\qquad \bar c=|n+2m|+1\,.
\end{gather*}
The characteristic exponents of this equation at $0$ and $1$ are
given respectively by $(0,-|n+2m|)$ and $(0,-|n-2m|)$.

It can be readily checked that $\tv$ belongs to the domain of
$\tT_{nm}$ if and only if $\hv$ is a bounded solution of
Eq.~\eqref{eqtw}, which implies~\cite[15.3.6]{AS70} that $\hv$ is a
polynomial. This only happens if $\bar a$ or $\bar b$ equals a
nonpositive integer $-j$, that is, when
\[
\La=2\Big(2j(j+1)+\big(|n+2m|+|n-2m|\big)(2j+1)+|n+2m||n-2m|+2m^2+n^2\Big)\,.
\]
In this case, $\hv(z)$ is a constant multiple of the Jacobi polynomial $P^{(|n+2m|,|n-2m|)}_j(1-2z)$. The spectral theorem
ensures that the corresponding eigenfunctions $\tv$ of $\tT_{nm}$ define
an orthogonal basis of $L^2((0,1))$, while their squared norm $\int_0^1|\tv(z)|^2\,dz$ can be readily shown to be~\cite[22.2.1]{AS70}
\begin{multline}\label{normv}
  \int_0^1z^{|n+2m|}(1-z)^{|n-2m|}\,P^{(|n+2m|,|n-2m|)}_j(1-2z)^2\,dz\\ =\frac{(j+|n+2m|)!\,(j+|n-2m|)!}{(2j+|n+2m|+|n-2m|+1)\,j!\,(j+|n+2m|+|n-2m|)!}\,.
\end{multline}
Since the change of variables
$\theta\mapsto z$ defines a unitary isomorphism
\[
L^2\big((0,\pi),\sin\theta\,d\theta\big)\ni v\mapsto \tv\in L^2\big((0,1),2\,dz\big)\,,
\]
the statement
of the lemma readily follows by inverting this transformation and normalizing the eigenfunctions using~\eqref{normv}.
\end{proof}

Let us now consider the differential operator
\begin{equation}\label{S}
  S_{ml}(\La):=\frac1{\rho(y)}\diff{y}\rho(y)\,w(y)\,r(y)\,\diff y -\frac1{w(y)}\bigg(\frac{\si l}\tau\bigg)^2- \frac9{r(y)}\bigg(2m-h(y)\,\frac{\si l}\tau\bigg)^2
-\frac{6\La}{1-y}\,,
\end{equation}
arising from~\eqref{Denml}, which depends on
a real parameter $\La\geq0$. It is clear that we cannot hope to
express the solutions of the formal eigenvalue equation
\begin{equation}\label{eqS}
  S_{ml}(\La)\,w=-\la w
\end{equation}
in closed form using special functions because~\eqref{eqS} is a
Fuchsian differential equation with {\em seven}\/ regular singular
points, located at the three roots of the cubic~\eqref{cubic}, at
$1$, at $± a^{1/2}$ and at infinity. However, the information
contained in the following lemma will suffice for our purposes:

\begin{lemma}\label{L.S}
For all $\La\geq0$, the differential operator~\eqref{S} defines a
nonnegative self-adjoint operator in $L^2((y_-,y_+),\rho(y)\,dy)$,
which we also denote by $S_{ml}(\La)$, whose domain consists of the
functions $w\in \AC((y_-,y_+))$ such that $S_{ml}(\La)w\in
L^2((y_-,y_+))$ and
  \begin{equation*}%\label{dom}
\lim_{y\searrow y_-}y\,w'(y)=0\; \text{ if }m=(2p-q)\si l/4 \quad\text{ and }\quad \lim_{y\nearrow y_+}y\,w'(y)=0\; \text{ if }m=-q\si l/4\,.
\end{equation*}
Its spectrum consists of a decreasing sequence of eigenvalues
$(-\ell_{mlk}(\La))_{k\in\NN}\searrow-\infty$ of multiplicity one
whose associated normalized eigenfunctions $w_{mlk}(\La)$ are
$\cO((y_+-y)^{|m+q\si l/4|})$ as $y\nearrow y_+$ and
$\cO((y-y_-)^{|m+(q-2p)\si l/4|})$ as $y\searrow y_-$.
\end{lemma}
\begin{proof}
Let $y_\ep$ be one of the endpoints of the interval $(y_-,y_+)$ and
set $\zeta:=y-y_\ep$. An easy computation shows that
  \[
a-3y^2+2y^3=-6y_\ep(1-y_\ep)\,\zeta+\cO(\zeta^2)\,,\qquad r(y)=-\frac\zeta{3y_\ep}+\cO(\zeta^2)
  \]
  as $y\to y_\ep$, which shows that the differential equation~\eqref{eqS} can be asymptotically written as
  \[
-\big(12y_\ep\,\zeta+\cO(\zeta^2)\big)\,\tw''(\zeta)-\big(12y_\ep+\cO(\zeta)\big)\, \tw'(\zeta)+\Bigg[ \frac{3y_\ep}\zeta\bigg(2m-h(y_\ep)\frac{\si l}\tau\bigg)^2+\cO(1)\Bigg]\,\tw(\zeta)=0\,,
  \]
with $\tw(\zeta):=w(\zeta+y_\ep)$ standing for the expression of the function $w(y)$ in the new variable~$\zeta$.

It then follows that the characteristic exponents of the
equation~\eqref{eqS} at $y_\ep$ are $±\nu_\ep$, with
$\nu_\ep:=|m-h(y_\ep)\si l/(2\tau)|$. Using~\eqref{pq}
and~\eqref{tausi}, one can immediately derive the more manageable
formula
  \begin{equation}\label{nuep}
\nu_+=\big| m+q\si l/4\big|\,,\qquad \nu_-=\big|m+(q-2p)\si l/4\big|\,.
\end{equation}
Since $\si$ is even by Lemma~\ref{L.Y}, it stems from the latter
equation that $2\,\nu_\ep$ is a nonnegative integer. Therefore, it
is standard that the symmetric operator defined by~\eqref{S} on
$C^\infty_0((y_-,y_+))$ is in the limit point case at $y_\ep$ if and
only if $\nu_\ep\neq0$. If $\nu_+\nu_-\neq0$, the latter operator is
then essentially self-adjoint on $C_0^\infty((y_-,y_+))$, and has a
unique self-adjoint extension of domain~\cite{DS88}
\begin{equation*}
\cD:=\big\{w\in \AC((y_-,y_+)):S_{ml}(\La)\,w\in L^2((y_-,y_+))\big\}\,..
\end{equation*}

When $\nu_+\nu_-=0$, the above symmetric operator is not essentially
self-adjoint. In this case, in order to rule out logarithmic
singularities we shall choose its Friedrichs extension~\cite{MZ00},
whose domain consists of the functions $w\in\cD$ such that
\[
\lim_{y\searrow y_-}y\,w'(y)=0\;\text{ if }\nu_-=0\quad\text{ and }\quad \lim_{y\nearrow y_+}y\,w'(y)=0\;\text{ if }\nu_+=0\,,
\]
It is well known~\cite{DS88} that $S_{ml}(\La)$ is then a
nonnegative operator with compact resolvent and that its eigenvalues
are nondegenerate.
\end{proof}
\begin{remark}\label{R.smooth}
It should be noticed that the critical exponents~\eqref{nuep} are
half-integers rather than integers, because in $|y-y_\ep|$ is
proportional to the square of the distance to the pole $y_\ep$, as
discussed in~\cite{GMSW04a}.
\end{remark}

Lemmas~\ref{L.T} and~\ref{L.S} provide us with all the information
we need in order to derive the following eigenfunction expansion for
the Laplacian, which is the main result of this section:

\begin{theorem}\label{T.spec}
Let $u_{nmlkj}:Y^{p,q}\to\CC$ be the analytic functions on $Y^{p,q}$
given by
  \begin{equation}\label{u}
u_{nmlkj}:=v_{nmj}(\theta)\,w_{mlk}(\La_{nmj})(y)\,\frac{\e^{\I(n\phi+2m\psi+\si l\al/\tau)}}{(2\pi)^{3/2}}
\end{equation}
and set $\la_{nmlkj}:=\ell_{mlk}(\La_{nmj})$. Then $\{u_{nmlkj}:j,k\in\NN,\; l,m,n\in\ZZ\}$ is an orthonormal basis of $L^2(Y^{p,q})$ and
  \begin{equation}\label{Deu}
\De u_{nmlkj}=-\la_{nmlkj}\,u_{nmlkj}\,.
\end{equation}
\end{theorem}
\begin{proof} We know that, by construction, $\{v_{nmj}\otimes w_{mlk}:j,k\in\NN\}$ is
a basis of $L^2((y_-,y_+),\rho(y)\,dy)\otimes L^2((0,\pi),
\sin\theta\,d\theta)$ for each $n,m,l\in\ZZ$. By Lemma~\ref{L.Y},
this implies that $\{u_{nmlkj}:j,k\in\NN,\; l,m,n\in\ZZ\}$ is then a
basis of $L^2(Y^{p,q})$. In the light of Lemma~\ref{R.smooth}, or
after a short computation, it is apparent that the functions
$u_{nmlkj}$ are analytic, and therefore lie in the domain
$H^2(Y^{p,q})$ of the Laplacian. From
Eq.~\eqref{DeDenml}--\eqref{Tnm} and~\eqref{S} and from
Lemmas~\ref{L.T} and~\ref{L.S} we subsequently infer that the
Laplace operator in diagonal in this basis and Eq.~\eqref{Deu}
holds, as claimed.
\end{proof}

\begin{remark}
It should be observed that the proof fails if one takes $v_{nmj}$ or
$w_{mlk}(\La)$ to be the eigenfunction basis corresponding to a
different self-adjoint extension of~\eqref{Tnm} or~\eqref{S}, since
in this case the above eigenfunctions will present logarithmic
singularities which will prevent some of the elements $u_{nmlkj}$ of
the basis of $L^2(Y^{p,q})$ from being in the domain of the
Laplacian. This is due to the fact that the action
of the Laplacian on these logarithmically divergent functions will
give rise to Dirac-type singularities.
\end{remark}

To conclude, some remarks on the form of the
eigenfunctions~\eqref{u} are in order. First, one should observe
that the angular dependence of the eigenfunctions (in $\al$ and
$\psi$) is quite different from, e.g., that of axisymmetric
eigenfunctions in Euclidean space. This is a consequence of the
considerable geometric complexity of the manifold $Y^{p,q}$ and its
fibration structure, and indeed one can gain some additional insight
into its geometry by fixing $y=y_±$ or
$(\theta,\phi)=(\theta_0,\phi_0)$ and studying the expression of the
eigenfunctions. From an analytic point of view, the main advantage
of Theorem~\ref{T.spec} is that most of the properties of the
Laplacian in $Y^{p,q}$ can be analyzed through the simpler
one-dimensional Sturm--Liouville operator $S_{ml}(\La)$.
Theorem~\ref{T.spec} will be crucial in our analysis of the wave
equation in $M$, which is the main result of this paper.

\section{The Klein--Gordon equation in $\AdS^5× Y^{p,q}$}
\label{S.wave}

We will denote by $\AdS^5$ the simply connected Lorentzian
5-manifold of constant sectional curvature $-\ka$, for fixed
$\ka>0$. It is well known that $\AdS^5$ is diffeomorphic to $\RR^5$.
If $\vartheta\equiv (\vartheta^1,\vartheta^2,\vartheta^3)\in[0,\pi]×
[0,\pi]×\SS$ are the standard coordinates on the $3$-sphere, the
metric on $\AdS^5$ can be globally written as
\[
g_\ka:= \frac{-dt^2+dx^2+\cos^2x\,g_{S^3}}{\ka\,\sin^2x}\,,
\]
where $t\in\RR$, $x\in (0,\pi/2]$ and
\[
g_{S^3}:=(d\vartheta^1)^2+\sin^2\vartheta^1\,(d\vartheta^2)^2+ \sin^2\vartheta^1\,\sin^2\vartheta^2\,(d\vartheta^3)^2
\]
is the canonical metric on $S^3$. The Laplacian and the Riemannian
volume measure on $S^3$ will be respectively denoted by $\De_{S^3}$
and
\[
d\om:=\sin^2\vartheta^1\,\sin\vartheta^2\,d\vartheta^1\,
d\vartheta^2\,d\vartheta^3\,.
\]

Let us now focus of the $10$-dimensional Lorentzian manifold
$\AdS^5× Y^{p,q}$, endowed with the metric $\bar g:=g_\ka\oplus g$.
It is apparent that there is no loss of generality in assuming that
the Ricci curvature of $Y^{p,q}$ is $4g$, as in
Section~\ref{S.Fourier}. Its associated wave operator will be
denoted by $\overline\square:=\bar
g^{ij}\,\overline\nabla_i\,\overline\nabla_j$, and we shall call
$\Si^T$ the spacelike hypersurface in $\AdS^5× Y^{p,q}$ defined by
$t=T$. We will occasionally use the above coordinates to naturally
identify each $\Si^T$ with a fixed time slice~$\Si$ and define the
measure $d\nu\,d\om\,d\mu$ on $\Si$ or $\Si^T$, with
$d\nu:=2\cot^3x\,dx$. Notice that $d\nu\,d\om\,d\mu$ is not the
hypersurface measure on $\Si^T$.

In this section we shall show that, under appropriate assumptions to
be made precise below, the Klein--Gordon equation in $\AdS^5×
Y^{p,q}$,
\begin{subequations}\label{Cauchy}
\begin{equation}\label{prob1}
\big(\overline\square- M^2)\vp=0
\end{equation}
with Cauchy data
\begin{equation}\label{prob2}
  \vp\big|_{\Si^0}=\vp^0\,,\qquad \frac{\pd\vp}{\pd t}\bigg|_{\Si^0}=\vp^1\,,
\end{equation}
\end{subequations}
admits a unique propagator, for any value of the constant $M\geq0$.
By a {\em propagator} we mean a bilinear map
\begin{equation*}%\label{cR}
\cR: C^\infty_0(\Si)× C^\infty_0(\Si)\to C^\infty(\AdS^5× Y^{p,q})\cap C^\infty\big(\RR,L^2(\Si,d\nu\,d\om\,d\mu)\big)
\end{equation*}
which maps each Cauchy data $(\vp^0,\vp^1)\in  C^\infty_0(\Si)×
C^\infty_0(\Si)$ to a smooth solution $\vp\equiv \cR(\vp^0,\vp^1)$
of the Cauchy problem~\eqref{Cauchy}.

The analysis of the Cauchy problem~\eqref{Cauchy} presents some
additional difficulties related to the fact that $\AdS^5× Y^{p,q}$
is not globally hyperbolic. Generally speaking, the existence or
uniqueness of global solutions to the Cauchy problem for the
Klein--Gordon equation is not granted on a manifold which fails to
be globally hyperbolic. The case of static, non-globally hyperbolic
manifolds such as $\AdS(\ka)× Y^{p,q}$ is quite special, however,
and propagators on these spaces can be analyzed using results of
Wald~\cite{Wa80} and Ishibashi and Wald~\cite{IW03,IW04}.

For various linear hyperbolic equations, Ishibashi and Wald discuss
the existence and uniqueness of propagators satisfying three
essential assumptions~\cite[Sect.\ 2]{IW03} which that are required
for the propagation to be physically sensible. Roughly speaking,
these conditions mean that the propagation in causal, invariant
under time translation and reflection, and that it preserves an
appropriate energy functional. For the reader's convenience, we
present these assumptions in the Appendix. Propagators for the
Cauchy problem~\eqref{Cauchy} which satisfy
Assumptions~\ref{A1}--\ref{A3} in the Appendix will be called {\em
admissible}. Our main result is Theorem~\ref{T.main}, where we prove
that the Cauchy problem~\eqref{Cauchy} has a unique admissible
propagator and derive a manageable integral spectral representation
for the solution.

Before stating this theorem, we need to introduce some further
notation. Let us consider the linear differential operators defined
by
\begin{equation}\label{L} 
  L(s,\la):=-\frac{\pd^2}{\pd x^2}+3\big(\tan x+\cot x\big)\,\diff{x}+\frac{s(s+2)}{\cos^2x}+\frac{M^2+\la}{\ka\,\sin^2 x}\,,
\end{equation}
where $s\in \NN$ and $\la\geq0$ are constants. We shall denote by
$\cL^0(s,\la)$ the positive symmetric operator in
$L^2((0,\frac\pi2),d\nu)$ with domain $C^\infty_0((0,\frac\pi2))$
defined by~\eqref{L}.

An orthonormal basis of $L^2(S^3,d\om)$ is given by the spherical harmonics (cf.\ e.g.~\cite{Sa78})
\begin{equation}\label{sph.harm}
  Y^{s_1s_2s_3}(\vartheta):=N_{s_1s_2s_3}\,
\sin^{s_2}\vartheta^1\, C_{s_1-s_2}^{(s_2+1)}(\cos\vartheta^1)\, P_{s_2}^{s_3}(\cos\vartheta^2)\,\e^{\I s_3\vartheta^3}\,,
\end{equation}
where $C_r^{(l)}$ and $P_l^m$ respectively denote the Gegenbauer
polynomials and the associated Legendre functions, $(s_1,s_2,s_3)\in
\NN^2×\ZZ$ with $s_1\geq s_2\geq|s_3|$, and the normalization
constants are
\[
N_{s_1s_2s_3}:=\bigg(\frac{2^{2s_2-1}(s_1+1)\,(2s_2+1)\,(s_1-s_2)!\,(s_2-s_3)!\,(s_2!)^2}{\pi^2(s_1+s_2+1)!\,(s_2+s_3)!}\bigg)^{1/2}
\]
The spherical harmonics satisfy the differential equation
\begin{equation}\label{S3}
  \De_{S^3}Y^{s_1s_2s_3}=-s_1(s_1+2)\, Y^{s_1s_2s_3}\,.
\end{equation}
For the ease of notation, we will consider the set of multi-indices
\[
B:=\big\{\be\equiv(\be_1,\dots,\be_8): \be_1,\be_2,\be_7,\be_8\in \NN,\; \be_3,\be_4\be_5,\be_6\in\ZZ,\; \be_1\geq \be_2\geq|\be_3|\big\}
\]
and denote by $\Psi_\be:S^3× Y^{p,q}\to\CC$ the smooth functions given by
\[
\Psi_\be(\vartheta,\eta):= Y^{\be_1\be_2\be_3}(\vartheta)\, u_{\be_4\be_5\be_6\be_7\be_8}(\eta)\,,
\]
where $u_{nmlkj}$ was defined in~\eqref{u}. We shall also use the notation
\begin{equation}\label{lambdac}
\la_\be:=\la_{\be_4\be_5\be_6\be_7\be_8}\quad \text{and}\quad c_\be:=\bigg(4+\frac{M^2+\la_\be}\ka\bigg)^{1/2}\,.
\end{equation}
A multi-index $\be$ should be thought of as an ordered $8$-uple
consisting of the spectral parameters $(s_1,s_2,s_3,n,m,l,k,j)$ used
in~\eqref{u} and~\eqref{sph.harm}.

\begin{lemma}\label{L.L}
For each $\be\in B$, the operator $\cL^0(\be_1,\la_\be)$ is
essentially self-adjoint. If $\cL_\be$ denotes its  closure and and
$F:[0,\infty)\to \CC$ is a bounded continuous function, then
\begin{equation}\label{F}
  F(\cL_\be)\,f(x)=\sum_{i\in\NN} F(\Om^\be_i)\, \bigg(\int_0^{\frac\pi2}f(x')\, \overline{f^\be_i(x')}\,d\nu(x')\bigg)\,f^\be_i(x)
\end{equation}
a.e.\ for all $f\in L^2((0,\frac\pi2),d\nu)$, where for each $i\in\NN$ the function
  \begin{equation}\label{fbei}
f^\be_i(x):=\bigg(\frac{2(2i+s+c_\be+2)\,i!\,\Ga(i+s+c_\be+2)}{(i+s+1)!\,\Ga(i+c_\be+1)}\bigg)^{1/2}\cos^{\be_1}x\,\sin^{2+c_\be}x\, P_i^{(\be_1+1,c_\be)}(-\cos2x)
\end{equation}
is a normalized eigenfunction of $\cL_\be$  with eigenvalue
\begin{equation}\label{varrho}
  \Om^\be_i:=\big(2i+s+c_\be+2\big)^2\,.
\end{equation}
\end{lemma}
\begin{proof} Let us consider the change of variables $\xi:=\cos^2x$, which maps
$L^2((0,\frac\pi2),d\nu)$ onto $L^2((0,1),\xi(1-\xi)^{-2}d\xi)$ and
transforms the ordinary differential equation $L(\be_1,\la_\be)f=\Om
f$ into
  \begin{equation}\label{Lxi}
4\xi(1-\xi)\,\tf''(\xi)+4(2-\xi)\,\tf'(\xi)+\bigg(\Om-\frac{\be_1(\be_1+2)}\xi- \frac{c_\be^2}{1-\xi}\bigg)\,\tf(\xi)=0\,,
\end{equation}
where $\tf$ stands for the expression of the function $f$ under the
above change of variables. This is a Fuchsian differential equation
with three regular singular points: $0$ (with characteristic
exponents $\be_1/2$ and $-1-\be_1/2$), $1$ (with characteristic
exponents $1+c_\be/2$ and $1-c_\be/2$) and infinity (with
characteristic exponents $±\Om^{1/2}/2$). From the expression for
the exponents and Eq.~\eqref{lambdac}, it is manifest that Eq.~\eqref{Lxi} does not admit any solutions in
$L^2((0,1),\xi(1-\xi)^{-2}d\xi)$ when $\imag\Om\neq0$, which implies
that the only self-adjoint extension of $\cL^0(\be_1,\la_\be)$ is
its closure, $\cL_\be$.

An easy computation shows that
$\hf(\xi):=\xi^{-\be_1/2}(1-\xi)^{-1-c_\be/2}\,\tf(\xi)$ satisfies
the hypergeometric equation
\begin{equation*}%\label{eqhf}
\xi(1-\xi)\,\hf''(\xi)+\big(\bar c-(\bar a+\bar b+1)\xi\big)\,\hf'(\xi)-\bar a\bar b\,\hf(\xi)=0\,,
\end{equation*}
with parameters
\begin{gather*}
  \bar a:=\frac{2+\be_1+c_\be-\Om^{1/2}}2\,,\qquad \bar b:=\frac{2+\be_1+c_\be+\Om^{1/2}}2\,, \qquad \bar c:=\be_1+2\,.
\end{gather*}
It is standard (cf.\ e.g.~\cite[15.3.6]{AS70}) that a solution $\tf$
to the above equation lies in $L^2((0,1),\xi(1-\xi)^{-2}d\xi)$ if
and only if $\hf$ is a polynomial. In turn, this is equivalent to
say that $\bar a=-i$ with $i\in\NN$, i.e., that
\[
\Om=\big(2i+s+c_\be+2\big)^2\,,
\]
which implies that $\hf(\xi)$ is proportional to the polynomial
$P_i^{(\be_1+1,c_\be)}(1-2\xi)$. When rewritten in terms of the
variable $x$, this readily yields the expressions~\eqref{fbei}
and~\eqref{varrho} for the eigenvalues and eigenfunctions of
$\cL_\be$, the normalization constant being easily read off from the
formula~\cite[22.2.1]{AS70}
\[
\int_0^1\xi^{\be_1+1}(1-\xi)^{c_\be}P_i^{(\be_1+1.c_\be)}(1-2\xi)^2\,d\xi=\frac{(i+\be_1+1)!\,\Ga(i+c_\be+1)}{2(2i +\be_1+c_\be+2)\,i!\,\Ga(i+\be_1+c_\be+2)}\,,
\]
with $\Ga$ being Euler's Gamma function. Eq.~\eqref{F} now follows through continuous functional calculus~\cite{DS88}.
\end{proof}

\begin{theorem}\label{T.main}
There exists a unique admissible propagator for the Cauchy problem~\eqref{Cauchy}, which is given by
  \begin{equation}\label{repr}
    \vp(t,x,\vartheta,\eta):=\sum_{\be\in B} \bigg(\cos\big(t\,\cL_\be^{1/2}\big)\,\vp^0_\be(x) +\frac{\sin(t\, \cL_\be^{1/2})}{\cL_\be^{1/2}}\,\vp^1_\be(x)\bigg)\,\Psi_\be(\vartheta,\eta)\,.
\end{equation}
Here
  \begin{equation*}%\label{vpjbe}
\vp^j_\be(x):=\int_{S^3× Y^{p,q}} \vp^j(x,\vartheta,\eta)\,\overline{\Psi_\be(\vartheta,\eta)}\,d\om(\vartheta)\,d\mu(\eta)\,,\qquad j=0,1,
\end{equation*}
and $\cos(t\,\cL_\be^{1/2})\, \vp^j_\be$ and $\cL_\be^{-1/2}\sin(t\, \cL_\be^{1/2})\,\vp^j_\be$ are defined through the formula~\eqref{F}.
\end{theorem}
\begin{proof}
The wave operator in $\AdS^5× Y^{p,q}$ reads
\begin{equation}\label{square}
\overline{\square}=\ka\bigg(-\sin^2x\,\frac{\pd^2}{\pd t^2}+\sin^2x\,\tan^3x\diff x \cot^3x\diff x+\tan^2x\,\De_{S^3}\bigg)+\De\,.
\end{equation}
Let us assume that $\cR$ is a propagator for the Cauchy
problem~\eqref{Cauchy} and set $\vp:=\cR(\vp^0,\vp^1)$, so that
$\vp|_{\Si^T}\in L^2(\Si,d\nu\,d\om\,d\mu)$ for almost every
$T\in\RR$. Let us decompose $\vp$ as
  \[
\vp(t,x,\vartheta,\eta)=\sum_{\be\in B}\vp_\be(t,x)\,\Psi_\be(\vartheta,\eta)\,,
\]
with
\[
\vp_\be(t,x):=\int_{S^3× Y^{p,q}} \vp(t,x,\vartheta,\eta)\,\overline{\Psi_\be(\vartheta,\eta)}\,d\om(\vartheta)\,d\mu(\eta)\,.
\]

From Eq.~\eqref{square} it then follows that $\vp_\be$ satisfies the
Cauchy problem
\begin{equation}\label{Cauchybe}
  \frac{\pd^2\vp_\be}{\pd t^2}+ L(\be_1,c_\be)\vp_\be=0\,,\qquad \vp_\be(0,\cdot)=\vp^0_\be\,,\qquad \frac{\pd\vp_\be}{\pd t}(0,\cdot)=\vp^1_\be\,,
\end{equation}
where, by hypothesis, $\vp_\be^j\in C^\infty_0((0,\frac\pi2])\subset
L^2((0,\frac\pi2),d\nu)$ for $j=0,1$. If we additionally assume that
$\vp_\be^j\in C^\infty_0((0,\frac\pi2))$, a theorem of Ishibashi and
Wald~\cite{IW03} ensures that the only admissible solutions
of~\eqref{Cauchybe} such that $\vp_\be(T,\cdot)\in
L^2(\Si,d\nu\,d\om\,d\mu)$ for a.e.\ $T\in\RR$ are given by
\begin{equation}\label{vpbe}
\vp_\be(t,x)=\cos\big(t\,\tcL_\be^{1/2}\big)\,\vp^0_\be(x) +\frac{\sin(t\, \tcL_\be^{1/2})}{\tL_\be^{1/2}}\,\vp^1_\be(x)\,,
\end{equation}
where $\tcL_\be$ is a positive self-adjoint extension of
$\cL^0(\be_1,\la_\be)$. As $\cL^0(\be_1,\la_\be)$ is essentially
self-adjoint by Lemma~\ref{L.L}, $\tcL_\be$ necessarily coincides
with the Friedrichs extension $\cL_\be$, which proves
Theorem~\ref{T.main} under the additional hypothesis that the
support of $(\vp^0,\vp^1)$ does not contain the point of $\Si^0$
given by $x=0$. To remove this hypothesis, it suffices to observe
that~\eqref{vpbe} still solves the Cauchy problem~\eqref{Cauchybe}
for arbitrary $\vp_\be^j\in C^\infty_0((0,\frac\pi2])$,
% because they belong to the domain of the operator $\cL_\be$,
% \[
% \cD_\be:=\big\{f\in\AC((0,\tfrac\pi2)): L(\be_1,\la_\be)f\in L^2((0,\tfrac\pi2),d\nu)\big\}\,,
% \]
thus ensuring the validity of~\eqref{repr} for arbitrary Cauchy data
$(\vp^0,\vp^1)\in C^\infty_0(\Si^0)× C^\infty_0(\Si^0)$, which is
what we had to prove.
\end{proof}

It is clear that Theorem~\ref{T.main} and its proof remain valid
when the Cauchy data are not assumed smooth and compactly
supported but taken in a Sobolev space of sufficiently high order,
but we shall not pursue this generalization here. It is also well
known that solutions to the inhomogeneous Cauchy problem can be
constructed through time integration of solutions to~\eqref{Cauchy}
using Duhamel's principle. In this case, the propagator is a
trilinear map
\[
\widehat\cR: C^\infty_0(\AdS^5× Y^{p,q})× C^\infty_0(\Si^0)× C^\infty_0(\Si^0)\to C^\infty(\AdS^5× Y^{p,q})\cap C^\infty\big(\RR,L^2(\Si,d\nu\,d\om\,d\mu)\big)
\]
mapping $(\Theta,\vp^0,\vp^1)$ to a solution $\Phi\equiv\widehat\cR(\Theta,\vp^0,\vp^1)$ of the PDE
\begin{equation}\label{CauchyPhi}
    \big(\overline\square-M^2\big)\Phi=\Theta\,, \qquad \Phi|_{\Si^0}=\vp^0\,,\qquad \frac{\pd\Phi}{\pd t}\bigg|_{\Si^0}=\vp^1\,.
  \end{equation}
For completeness we state the analog of Theorem~\ref{T.main} for the
inhomogeneous equation, which follows using the same reasoning as in
Theorem~\ref{T.main}. It should be noticed that the admissibility
conditions presented in the Appendix can be readily extended to the
case of inhomogeneous equations, mutatis mutandis.

\begin{corollary}\label{C.inhom}
There exists a unique admissible propagator $\widehat\cR$ for the
inhomogeneous Cauchy problem~\eqref{CauchyPhi}. The propagator is
given by $\Phi\equiv\widehat\cR(\Theta,\vp^0,\vp^1)=\vp+\tvp$, with
$\vp$ as in~\eqref{repr} and $\tvp$ defined by
  \begin{equation}\label{repr2}
\tvp(t,x,\vartheta,\eta):=\sum_{\be\in B} \bigg(\int_0^t\frac{\sin[(t-T)\cL_\be^{1/2}]}{\cL_\be^{1/2}}\Theta_\be(T,x)\,dT\bigg)\, \Psi_\be(\vartheta,\eta)\,,
\end{equation}
where
  \begin{equation*}%\label{vpjbe}
\Theta_\be(t,x):=\int_{S^3× Y^{p,q}} \Theta(t,x,\vartheta,\eta)\,\overline{\Psi_\be(\vartheta,\eta)}\,d\om(\vartheta)\,d\mu(\eta)
\end{equation*}
and the integral in~\eqref{repr2} is defined through the formula~\eqref{F}.
\end{corollary}

\section*{Appendix}

In this Appendix we will state and briefly discuss Ishibashi and
Wald's assumptions~\cite{IW03}, which must be satisfied by any
physically meaningful propagator $\cR$ of the Klein--Gordon
equation~\eqref{Cauchy}. To this end, we will denote by
$\vp\equiv\cR(\vp^0,\vp^1): \AdS^5× Y^{p,q}\to\CC$ the solution of
the Cauchy problem~\eqref{Cauchy} determined by $\cR$.

Let us start by defining the time translation and reflection
operator on $C^\infty(\AdS^5× Y^{p,q})$:
\[
(\cT_T\Phi)(t,x,\theta,\eta):=\Phi(t+T,x,\theta,\eta)\,,\qquad (\cP\Phi)(t,x,\theta,\eta):=\Phi(-t,x,\theta,\eta)\,,
\]
with $T\in\RR$. The causal future (resp.\ past) of a set $U\subset
\AdS^5× Y^{p,q}$ will be denoted by $J^+(U)$ (resp.\ $J^-(U)$). The
first condition we must impose on the propagator is that a solution
$\vp$ of~\eqref{Cauchy} must be compatible with causality:

\begin{assumption}\label{A1}
  The support of $\vp$ is contained in $J^+(\supp(|\vp^0|+|\vp^1|)\cup J^-(\supp(|\vp^0|+|\vp^1|))$.
\end{assumption}

The second assumption is that the propagation is compatible with the
time translation and reflection symmetries. In order to state this
condition, we need to recall~\cite[Lemma 2.1]{IW03} that
Assumption~\ref{A1} ensures that for any initial conditions
$(\vp^0,\vp^1)\in C^\infty_0(\Si^0)× C^\infty_0(\Si^0)$ there exists
some $\ep(\vp^0,\vp^1)>0$ such that the function
$\vp_T(x,\vartheta,\eta):=\vp(T,x,\vartheta,\eta)$ is smooth and
compactly supported for all $|T|<\ep(\vp^0,\vp^1)$. (Of course, the
function $\vp_T$ would be compactly supported for all $T$ in a
globally hyperbolic manifold.)

\begin{assumption}\label{A2}
Let $\vp$ be the solution associated with the Cauchy data
$(\vp^0,\vp^1)$ and $|T|\leq \ep(\vp^0,\vp^1)$. Then the solution
associated with the Cauchy data $(\vp_T,(\pd\vp/\pd t)_T)$ (resp.\
$(\vp^0,-\vp^1)$) is $\cT_{-T}\vp$ (resp.\ $\cP\vp$).
\end{assumption}

To state the third assumption, let us introduce the vector space
\[
\cV:=\big\{\Phi\equiv\sum_{i=1}^N\cT_{T_i}\cR(\vp^0_i,\vp^1_i): N\in\NN,\; T_i\in\RR,\; \,, (\vp^0_i,\vp^1_i) \in C^\infty_0(\Si^0)× C^\infty_0(\Si^0)\big\}
\]
of all finite linear combinations of solutions of the form
$\cT_{T}\cR(\vp^0,\vp^1)$. In the case of a globally hyperbolic
manifold, this would simply be the space $\cR(C^\infty_0(\Si^0)×
C^\infty_0(\Si^0))$ of solutions to the Cauchy
problem~\eqref{Cauchy} with data in $C^\infty_0$.

Assumption~\ref{A3} consists of three related conditions ensuring
the existence of a well defined conserved energy functional $E$. The
first one asserts that this energy is invariant under time
translations and reflections. By Eq.~\eqref{square}, the
equation~\eqref{prob1} can be rewritten as
\begin{equation}\label{rewrite}
\bigg(-\frac{\pd^2}{\pd t^2}+\tan^3x\diff x \cot^3x\diff x+\frac{\De_{S^3}}{\cos^2x}+\frac{\De-M^2}{\ka\,\sin^2x}\bigg)\vp=0\,,
\end{equation}
so the second condition guarantees that the $E$ reduces to the
`naive' energy functional for Eq.~\eqref{rewrite} in certain cases.
Finally, the third condition requires that the topology defined by
the energy functional be compatible with the weak $C^\infty$
topology on the Cauchy data.

\begin{assumption}\label{A3}
There exists an inner product $E:\cV×\cV\to\CC$ such that the
following properties hold:
\begin{enumerate}
  \item For all $\Phi_1,\Phi_2\in\cV$ and $T\in\RR$,
    \[
    E(\Phi_1,\Phi_2)= E(\cP\Phi_1,\cP\Phi_2)= E(\cT_T\Phi_1,\cT_T\Phi_2)\,.
    \]
  \item For all $\Phi\in\cV$,
    \[
E(\Phi,\vp)=\int_{\Si^0}\bigg[\overline{\frac{\pd \Phi}{\pd
t}}\,\vp^1-\overline{\Phi}\,\bigg(\frac{\pd^2}{\pd x^2} +3\big(\tan
x+\cot x\big)\,\diff
x+\frac{\De_{S^3}}{\sin^2x}+\frac{\De-M^2}{\ka\,\sin^2
x}\bigg)\vp^0\bigg]\,d\nu\,d\om\,d\mu\,.
\]
\item Let $(\Phi_n)_{n\in\NN}\subset \cV$ be a Cauchy sequence with respect
to the norm $\|\cdot\|_E$ defined by $E$ and suppose that there
exists $\Phi\in\cV$ such that the sequences
$(\Phi_n|_{\Si^0})_{n\in\NN}$ and $(\pd\Phi_n/\pd
t|_{\Si^0})_{n\in\NN}$ respectively tend to $\Phi|_{\Si^0}$ and
$\pd\Phi/\pd t|_{\Si^0}$ in the weak $C^\infty(\Si^0)$ topology.
Then
    \[
\lim_{n\to\infty}\big\|\Phi_n-\Phi\big\|_E=0\,.
    \]
  \end{enumerate}
\end{assumption}

\section*{Acknowledgments}

The authors are indebted to Keshav Dasgupta and James Sparks for
valuable discussions. A.E.\ is financially supported by a MICINN
postdoctoral fellowship and thanks McGill University for hospitality
and support. A.E.'s research is supported in part by the MICINN and
the UCM--Banco Santander under grants no.~FIS2008-00209
and~GR58/08-910556. The research of N.K. is supported by NSERC grant
RGPIN 105490-2004.

%\small

\end{document}